\documentstyle[prl,aps,multicol,psfig,epsfig]{revtex}

\begin{document}
\draft

\author{V.\ M.\ Pudalov$^{a,b}$, M.\ E.\ Gershenson$^{a}$, and H.\ Kojima$^{a}$}
\address{$^a$ Department of Physics and Astronomy, Rutgers University,
New Jersey 08854, USA}
\address{$^b$ P.\ N.\ Lebedev Physics
Institute/Lebedev Research Centre in Physics, 119991 Moscow, Russia}

\date{\today}
\title{Absence of Ferromagnetic Instability
at the Metal-Insulator Transition in Si-inversion Layers}
\maketitle

\begin{abstract}
We have measured the
Shubnikov-de Haas oscillations in high-mobility Si MOS structures
over a wide range of the carrier densities $n\geq 0.77\times
10^{11}$cm$^{-2}$. This range includes the critical density $n_c$
of the metal-insulator transition for two samples studied. The
periodicity of oscillations clearly demonstrates that the electron
states remain fourfold degenerate down to and at the 2D MIT. Both
the effective  spin susceptibility $\chi^*$ and mass $m^*$
remain finite and show no signatures of divergency
at the critical density for both samples  studied. To
test possible divergency of  $\chi^*(n)$ and
 $m^*(n)$ at even lower densities, we have analyzed the
data on $\chi^*(n)$ and $m^*(n)$ in terms of a critical dependence
$\chi^*, m^* \propto (n/n_0 -1)^{-\alpha}$. Our data suggest that
$\chi^*$ and $m^*$ may diverge at $n_0 \lesssim 0.5\times 10^{11}$cm$^{-2}$
($r_s\geq 12$), which is significantly  smaller than $n_c$.

\end{abstract}


\vspace{-0.1in}
\begin{multicols}{2}

Despite intensive experimental and theoretical efforts (see, e.g.,
Ref.~\cite{aks} for a  bibliography),
the origin  of the apparent
``metal-insulator transition in two dimensions'' (2D MIT) remains to
be the subject of ongoing discussion. This phenomenon addresses a
fundamental problem of the ground state of strongly correlated and
disordered electron systems. In (100) Si inversion layers, the 2D
MIT is observed at a sample-dependent  critical electron density
$n_c \sim 1\times 10^{11}$\,cm$^{-2}$ \cite{JETPL_98a}.

One of the important unsolved problems is a possible magnetic
instability \cite{finkelstein} in spin  or valley systems. The
electron-electron interactions drive a 2D
 system towards magnetic transition;
numerical calculations for the critical value of $r_s$
\cite{note1} at the instability vary from 13 to 20
\cite{isihara_82}. According to these calculations, the
ferromagnetic transition is likely to be of first order with a
complete rather than a partial (ferrimagnetic) spin polarization.
An interesting interpretation
\cite{shashkin_0007402,vitkalov_0009454} of the parallel-field
magnetoresistance in Si inversion layers suggested a
ferromagnetic instability at or very close to $n_c$ [where $n_c
\approx (0.8-0.85)\times10^{11}$cm$^{-2}$, which corresponds to
$r_s \approx 9$]. The idea of magnetic instability and  its
possible link to the 2D MIT is important and requires a careful
examination.

In this Letter, we report on our experimental test of two possible
scenarios for the magnetic
instability: (i) complete spin and/or valley polarization occurs
spontaneously at a sample-dependent critical density of the MIT,
and (ii) $\chi^*$ diverges at a universal (sample-independent)
value of $n=n_0$. To test the first scenario, we measured the
Shubnikov-de Haas (SdH) oscillations and determined the degeneracy of
the electron system across the 2D MIT in two samples (down to
$n=0.77\times 10^{11}$cm$^{-2}$). We found that the period of
oscillations corresponds to the double-degenerate spin and valley
states even in the presence of external magnetic field $B\approx 0.5$\,T.
This rules out the possibility of a complete
spin/valley polarization at the 2D MIT. To test the
second scenario, we analyzed independent measurements of
 $\chi^*(n)$ and  $m^*(n)$, and found
that each could  be described by the same critical dependence
$(n/n_0-1)^{-\alpha}$, if we impose an upper limit on the density
of the instability, $n_0 \lesssim 0.5\times
10^{11}$cm$^{-2}$ ($r_s\geq 12$), and a lower limit on the
critical index $\alpha \gtrsim 0.6$. The density $n_0$ is
significantly lower than $n_c$ for the samples studied.

The measurements were performed on two Si-MOS samples: Si6-14
(peak mobility $\mu^{\rm peak}\simeq 2.2$m$^2$/Vs) and Si5
($4.3$m$^2$/Vs), with the critical density $n_c$ of
the apparent MIT $1.0\times10^{11}$cm$^{-2}$ and
$0.77\times10^{11}$cm$^{-2}$, respectively.
The MOS structures are made on a (001)-Si wafer
with [100] source-drain orientation; the gate oxide
thickness was $\approx 190$\,nm. The density of  electrons
was controlled by the gate voltage $V_g$ and determined from the
period of SdH oscillations. An in-plane field $B_{\parallel}\geq
0.02$\,T was applied for quenching the superconductivity of the Al
contact pads and gate electrode. Details of the experimental technique
can be found in Refs.~\cite{gm,crossed}

Typical SdH oscillations of the resisitivity $\rho_{xx}\equiv \rho$
are shown in Fig.~1\,a as a
function of $B_\perp$. Due to a high electron mobility, oscillations
were detected  in fields down to 0.14\,T. To examine directly the
first scenario, we focus on the  period of
the SdH oscillations, the quantity which is not renormalized by
interactions.

For such low densities as presented in Fig.~1, the oscillations
$\rho(B_\perp)$ are ``shaped'' mostly by the spin energy gaps
\cite{okamoto,gm,termination94,kravSSC2000}. Figure~1\,b shows
that the magnitude of oscillations increases with an in-plane
magnetic field $B_{\parallel}$. This confirms that the ratio of
the Zeeman energy $E_Z=g^*\mu_B B_\perp$ to the cyclotron energy
$\hbar \omega_c$ is within the interval $1/2 <
E_Z/\hbar\omega_c < 1$, in good agreement with the measured
values of $\chi^*(n)$ \cite{gm}, which control the calculated
energy spectrum (the upper inset to Fig.~1\,b). Curve {\em 5} in
Fig.~1\,a corresponds to the density $n=n_c$ for sample Si6-14.
The latter has been
determined in the insulating regime, $\rho
\propto \exp(-\Delta/k_B T)$,  by extrapolating the density dependence of the
activation energy $\Delta(n)$  to zero  \cite{prl93} (see the inset to Fig.~1\,b).

\vspace{0.05in}
\begin{figure}
\centerline{\psfig{figure=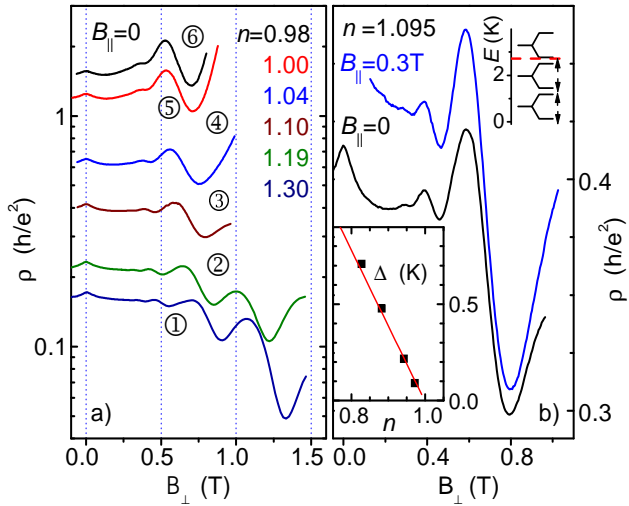,width=235pt,height=190pt
}}
\vspace{0.1in}
\begin{minipage}{3.2in}
\caption{a) SdH oscillations for  the sample Si6-14 at six
densities near $n_c$,  $T=0.2$\,K. Curves $3-6$ are terminated
at the onset of a large insulating peak in $\rho$
\protect\cite{pud_92}. b) Enhancement of oscillations with
$B_{\parallel}$. The upper inset shows the energy spectrum for
$B_\perp=0.5$\,T, $n=1.0\times 10^{11}$cm$^{-2}$, $g^*m^*/2m_b=4.35$,
$m^*=0.5m_e$ \protect\cite{gm}. Vertical arrows depict spin
polarization and the direction of the corresponding level shift
with $B_{\parallel}$. The lower inset illustrates
determination of the critical density
for Si6-14.
Densities are given in units of
$10^{11}$\,cm$^{-2}$.}
\end{minipage}
\label{fig1}
\end{figure}

In order to emphasize the low-field region, and to clearly
illustrate the SdH periodicity, we present the experimental data
normalized by the amplitude of the first SdH harmonic $A_1(B_\perp)$ \cite{SdH,gm}.
In evaluating $A_1$, we used the values of $\chi^*(n) \propto
g^*m^*$ and $m^*(n)$ measured in Ref.~\cite{gm}; the Dingle
temperature was adjusted to match damping of the measured oscillations.
Figure~2 shows oscillations of the resistivity $\delta
\rho/ \rho_{0}A_1$ as a function of the Landau level filling,
$\nu=nh/eB_\perp$. The $\delta \rho(\nu)/\rho_{0}A_1$ data (dots)
in Figs.~2\,a--c correspond to the $\rho(B_\perp)$ data {\em 1, 5} and {\em
6} in Fig.~1\,a.
It is important to limit the field range $B_\perp \leq
1T$ in the
analysis of the SdH oscillations  in order to diminish the magnetic-field-induced spin
polarization and reentrant quantum Hall effect-to-insulator
transitions \cite{pud_92}. The former limitation  was violated at
$\nu <5$: doubling of the period for $\nu=4$ in Fig.~2\,a
illustrates lifting  the spin degeneracy  by the perpendicular
field $B_\perp \sim 1.3\,T$. The latter limitation was violated for
$\nu<10$ in Figs.~2\,b-e and  may account for the large oscillation
amplitude.

For all the densities studied, including $n = n_c$ (Figs. 2\,b\,e), the
low-field oscillations $\delta \rho/ \rho_{0}$ have the
period $\Delta\nu = 4$ that corresponds to a double-degenerate
{(\em i.e. unpolarized)} spin and valley system. The minima of
$\delta \rho$ in Figs.~2 are located at $\nu=6, 10, 14, 18$,
in contrast to  $\nu =4, 8, 12, 16$, as observed for higher
densities. This is in agreement with earlier results
\cite{termination94,kravSSC2000,pud_92} and with the measured
$\chi^*$ values \cite{gm}. The sign of oscillations changes
 due to the Zeeman factor
$cos(\pi E_Z/\hbar\omega_c)$,  when $E_Z$ exceeds $\hbar\omega_c/2$
(at $r_s > 6.3$) \cite{gm,SdH}.

\vspace{0.05in}
\begin{figure}
\centerline{\psfig{figure=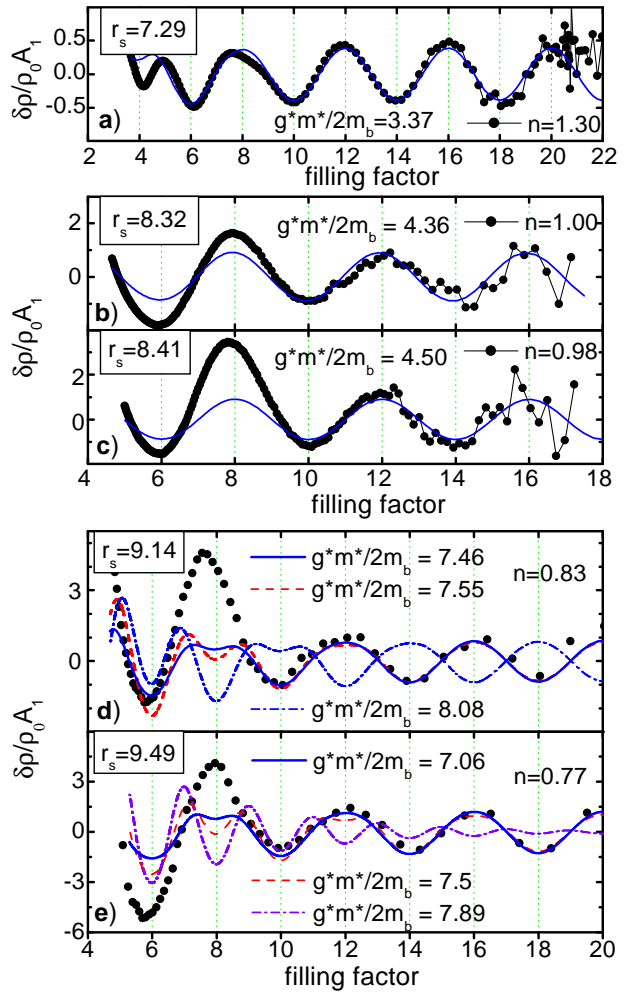,width=230pt}}
\vspace{0.1in}
\begin{minipage}{3.2in}
\caption{Oscillatory component of
the resistivity
measured for samples Si6-14
(panels a-c) and Si5
(panels d, e). The data are shown as dots, the
fits as lines \protect\cite{gm}; both are normalized by $A_1(B_\perp)$.
The temperature is 0.2\,K for traces (a - c) and 0.03\,K for
traces d and e.
The values of $n$ (in units of $10^{11}$\,cm$^{-2}$),
 $r_s$, and $g^*m^*$ are shown in each panel.}
\end{minipage}
\label{Fig.2}
\end{figure}

From these  data, we can estimate the spontaneous spin
polarization $P_0$  near  $n_c$.
The total spin polarization of the interacting
2D electron system is:
\begin{equation}
P  \equiv \frac{n_{\uparrow}-n_{\downarrow}}{n} = P_Z +P_0=
\frac{\chi^* B_\perp}{2 \mu_B n} +P_0.
\end{equation}
Figures~2\,c and 2\,e show
that the complete spin (or valley) polarization does not occur at least down to
$\nu=6$ ($P_Z\approx 0.3$), for both $B_\parallel =0$ and
$B_\parallel=0.3$T (Fig.~1b); thus, the spontaneous
component $P_0 <1-0.3=0.7$.  We can impose even  more restrictive
upper limit on the spontaneous spin polarization across the 2D
MIT, $P_0 \lesssim 0.18$, by noting that no nodes of beating
are seen in SdH oscillations in
Figs.~2\,c, 2\,e over the  interval of $\nu = 5 - 20$.
For example, a larger value of $P_0= 0.19$ would induce a beating node
in the SdH oscillations at $\nu=15$.
For even lower densities, the period
of the $\rho$ oscillations \cite{termination94,pud_92} demonstrates that
both the spin and valley states remain double-degenerate across
the 2D MIT.
These {\em direct} data provide very strong evidence against
 a complete spontaneous spin/valley polarization
at the sample-dependent critical density for $n \geq 0.77\times 10^{11}$cm$^{-2}$.

Below, we  explore the second scenario, a divergence of
 $\chi^*$ and  $m^*$
at a sample-independent density
$n_0$. Indeed, it has been found in  recent measurements
\cite{okamoto,gm} that  both quantities
increase with decreasing carrier concentration and are sample-independent
($\chi^*$ - to within $\pm 2\%$, and $m^*$ - $\pm 4$\%).
We examine our data on $\chi^*(n)$ and $m^*(n)$ \cite{gm,twomasses}
for sample Si6-14
 for possible critical density dependence in the form
$(n/n_0-1)^{-\alpha}$.
For a given $n_0$, the exponents $\alpha_\chi$ and $\alpha_m$
have been obtained from fitting
 $\chi^*(n)$ and $m^*(n)$. An example with
$n_0=0.53\times 10^{11}$cm$^{-2}$ and $\alpha=0.63$
is shown in  Figs.~3\,a,\,b. The exponents $\alpha_\chi$ and $\alpha_m$
versus $n_0$ (hash marks indicate the error bars) are
shown  in the inset in Fig.~3a.
The standard deviation of $\alpha(n_0)$  has such a shallow minimum
that the optimal $n_0$ value
could not be determined reliably;
the large uncertainty was mainly caused by
unknown critical range of densities.
Similar uncertain situation was encountered in the critical analysis
of the $m^*$ data for sample Si6-14 (see Figs.~3b and 3c).

We now include  into consideration
the data for sample Si5 which has
substantially lower $n_c$.  Three additional $m^*(n)$ points from Si5
are shown as diamonds in Fig.~3\,c. We find that the critical dependence cannot fit the data
when $n_0 $ is taken greater than  $0.65\times 10^{11}$cm$^{-2}$; this value sets
 the upper limit for possible $n_0$. Although only the $m^*(n)$
 data were available for Si5, $\chi^*(n)$ can be
estimated independently from the lineshape and phase of the SdH oscillations.
The SdH pattern is very sensitive to the ratio
$E_Z/\hbar\omega_c$: when this ratio becomes greater than 3/2 ($g^*m^*\geq 7.89$)
or smaller than 1/2 ($g^*m^*\leq 2.63$), the phase of oscillations
changes by $\pi$. Theoretical curves in Figs.~2\,d and 2\,e show that, even
before the phase reverses for all oscillations, the highest-field oscillation ($\nu=6$)
splits starting from $g^*m^* \approx 7$.
The absence of  such behavior in the measured $\delta\rho(B_\perp)$
traces enables us to obtain,
correspondingly, the upper and lower estimates for $g^*m^*$  at six densities in the range
from $n=0.768$ to $0.884\times 10^{11}$cm$^{-2}$.
Three of these estimates are shown by
vertical bars in Fig.~3a (three others, located between the shown bars,
are omitted for clarity).

Small values of $n_0 \ll n_c$ can be certainly accommodated by the critical dependence;
however, we searched for the upper limit on $n_0$, in order to determine
how close  $n_0$ could be to $n_c$. For this reason, we have used the {\em upper}
limits for $\chi^*$ (the top of bars in Fig.~3\,a) when we plotted the critical
dependence in Fig.~3\,a.
The $\chi^*(n)$ data  for both Si5 and Si6-14 obey a
common critical dependence only if
 we choose $n_0 \lesssim 0.53\times 10^{11}$cm$^{-2}$.
 This choice of $n_0$ also provides the lower
limit for the critical index $\alpha_\chi \geq 0.63$.
This procedure defines the range of densities
where the critical behavior holds:
e.g., for $n_0=0.53\times 10^{11}$cm$^{-2}$, this range corresponds to
$n <1.5\times 10^{11}$cm$^{-2}$.
 A similar but less restrictive conclusion follows  from the
critical analysis for $m^*$ (Figs.~3b,c): $n_0 \lesssim 0.65\times
10^{11}$cm$^{-2}$ and $\alpha_m \geq 0.4$.
It is important to note that the upper limit
on $n_0$ is  a factor of 1.5--2 lower than the critical density
$n_c$ for the samples studied.

\vspace{0.05in}
\begin{figure}
\centerline{\psfig{figure=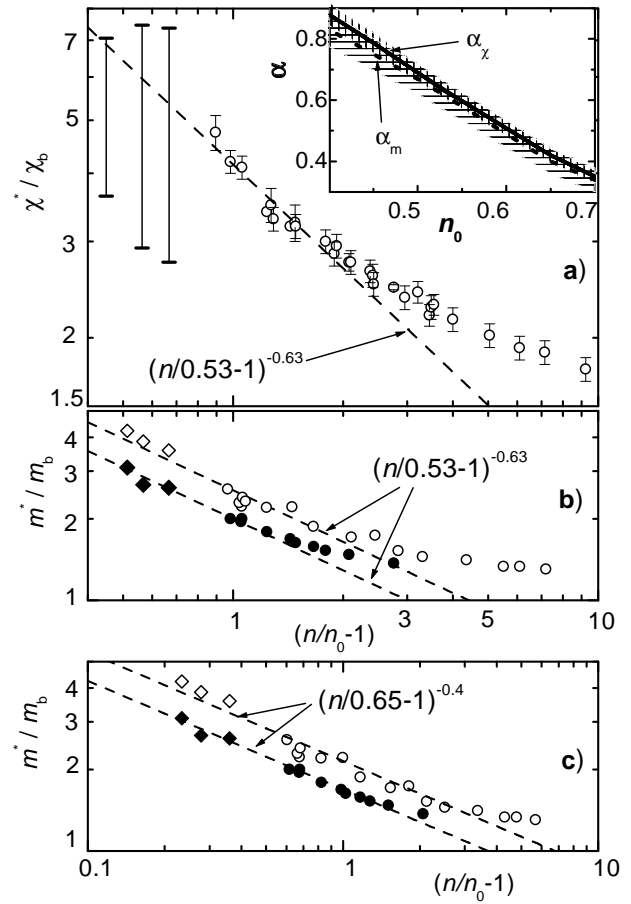,width=230pt}}
\vspace{0.1in}
\vspace{0.05in}
\begin{minipage}{3.2in}
\caption{ Log-log plots of (a) the spin susceptibility
$\chi^*/\chi_b$  and (b,c) the  mass $m^*/m_b$  vs $(n/n_0
-1)$: dots for sample Si6-14,  diamonds and bars for  Si5. The $m^*(n)$ data
are plotted for two values of $n_0 = 0.53$  and $0.65\times
10^{11}$\,cm$^{-2}$. On panels (a):
the vertical bars extend from the upper to lower   limits
for $\chi^*$, as discussed in the text.
On panels (b,c): open and closed symbols depict the upper
and lower estimate for $m^*$, obtained from the $T$-dependence
of SdH amplitude \protect\cite{gm,twomasses}. The
dashed lines show the critical behavior $\chi^*, m^* \propto
(n/n_0-1)^{-\alpha}$. The inset shows the critical indices
$\alpha_\chi$ and $\alpha_m$ vs $n_0$. }
\label{fig3}
\end{minipage}
\end{figure}

For a given $n_0$ value, the estimated
critical indices  $\alpha_\chi$ and $\alpha_m$ are close to each other
(see the inset to Fig.~3a).
This might be expected:
 there is no experimental indication for a
 critical behavior of $g^*(n)$ \cite{gm,SKDK0111478},
therefore $\chi^* \propto g^*m^*$ and $m^*$ should exhibit the same
$n$- dependence (a critical behavior or otherwise).

The conclusion on the absence of the magnetic instability at
$n\approx n_c$, which we have drawn from our analysis, differs from the
one suggested in Refs.~\cite{shashkin_0007402,vitkalov_0009454}.
There might be several reasons for this disagreement. Firstly, we
believe that the magnetoresistance data for in-plane fields,
analyzed in Refs.~\cite{shashkin_0007402,vitkalov_0009454}, might be indirectly
related to the spin susceptibility of {\em mobile} electrons.
Secondly, measurements in
Refs.~\cite{shashkin_0007402,vitkalov_0009454} were taken in {\em
strong} in-plane fields. The strong fields drive a 2D system into
the hopping regime \cite{shashkin}; the characteristic values of
$B_\parallel$ go to zero as
$n$ approaches $n_c$ \cite{aniso}. Moreover, even moderate  fields
$B_\parallel < E_F/g^*\mu_B$
 induce non-linearity of magnetization,  i.e. the $\chi^*(B)$-dependence
 \cite{gm,nonlinear}. In contrast, our
{\em direct} measurements of $\chi^*$ have been performed
in the {\em low-field} linear regime.

To summarize, we measured  Shubnikov-de Haas oscillations in weak
perpendicular fields  over a wide density range $n \geq 0.77 \times
10^{11}$cm$^{-2}$, which includes the sample-specific critical densities $n_c$ of the 2D
MIT for two different samples. It has been found that the period
of oscillations corresponds to the fourfold degeneracy of
spin/valley systems on both sides of the 2D MIT.
Our results demonstrate that the apparent 2D MIT
is not accompanied by
a spontaneous complete polarization of spins or valleys at zero
field. Moreover, the experimental data allow us to put an upper
limit $P_0<0.18$ on the value of a possible spontaneous
polarization at the transition.
We also
explored a possibility of critical behavior of
the renormalized spin susceptibility and the effective mass at a
sample-independent density $n_0$.
We found that the
divergence of both $\chi^*$ and $m^*$ is unlikely for
$n_0 > 0.65\times 10^{11}$cm$^{-2}$.
However, it may occur at lower densities
(significantly less than $n_c$): e.g. for $\chi^*$ -- at
$n_0\lesssim 0.5\times 10^{11}$cm$^{-2}$
and $\alpha \gtrsim 0.6$.

Authors are grateful to E.\ Abrahams, B.\ L.\ Altshuler, and D.\ L.\
Maslov for discussions. The work was supported by the NSF, ARO
MURI, NWO, NATO, RFBR, INTAS, and the Russian programs ``Physics of
Nanostructures'', ``Quantum and Non-linear Processes'',
``Quantum computing and telecommunications'',
``Integration of High Education and Academic Research'', and
``The State Support of Leading Scientific Schools''.

\end{multicols}

\begin{references}
\bibitem{note1} The parameter
$r_s = 2.63\times (10^{12}{\rm cm}^{-2}/n)^{1/2}$
which characterizes the electron-electron interactions,
is the  ratio of the Coulomb interaction energy $E_{ee}$ to the Fermi
energy $E_F$. The bare values of the Fermi energy $E_F$, the effective mass $m_b=0.19$,
and the dielectric constant $\kappa =7.8$ at the
Si-SiO$_2$ interface are used in its definition.
\bibitem{twomasses}The two values for the effective mass (open and closed symbols)
are its upper and lower border \protect\cite{gm} and
originate from lacking of adequate theory for the temperature
dependence of $T_D$ and $m^*$ under conditions of $k_Fl \sim 1$.
\end{references}
\end{document}